\documentclass[prb,twocolumn,showpacs,aps,superscriptaddress,floatfix]{revtex4}
\usepackage{amsmath}
\usepackage{amssymb}
\usepackage{bm}
\usepackage{graphicx}
\usepackage{color}
\usepackage{hyperref}
\usepackage{verbatim}

\begin{document}
\title{Surface charge conductivity of topological insulator in a magnetic field: effect of hexagonal warping}
\author{R.S. Akzyanov}
\affiliation{Moscow Institute of Physics and Technology, Dolgoprudny,
Moscow Region, 141700 Russia}
\affiliation{Dukhov Research Institute of Automatics, Moscow, 127055 Russia}
\author{A.L. Rakhmanov}
\affiliation{Moscow Institute of Physics and Technology, Dolgoprudny,
Moscow Region, 141700 Russia}
\affiliation{Dukhov Research Institute of Automatics, Moscow, 127055 Russia}
\affiliation{Institute for Theoretical and Applied Electrodynamics, Russian
Academy of Sciences, Moscow, 125412 Russia}

\begin{abstract}
We investigate the influence of the hexagonal warping on the transport properties of the topological insulators. We study the charge conductivity within Kubo formalism in the first Born approximation using low energy expansion of the Hamiltonian near the Dirac point. The effects of disorder, magnetic field and chemical potential value are analyzed in details. We found that the presence of the hexagonal warping effects significantly the conductivity of the topological insulator. In particular, it gives rise to the growth of the longitudinal conductivity with the increase of the disorder and anisotropic anomalous in-plane magnetoresistance. The hexagonal warping also affects the quantum anomalous Hall effect and anomalous out-of-plane magnetoresistance. The obtained results are consistent with the experimental data.
\end{abstract}


\maketitle

\section{Introduction}

Discovery of topological insulators (TIs) led to the exploding growth of interest to the topologically protected states~\cite{Konig2007,Zhang2009}. Topologically non-trivial band structure of the TI results in the existence of the conducting surface states, while the bulk of the sample is insulating. Such states has a helical Dirac cone dispersion and are protected against elastic backscattering~\cite{Hasan2010,Qi2011}. Robustness of such topologically protected states may be of interest for applications in the modern devices starting from the ultralow-dissipative memory~\cite{Fan2014} up to topological quantum computers~\cite{Fu2008}.

Transport properties of the TIs are in some aspects similar to graphene, the another Dirac material~\cite{CastroNeto2009}. The conductivity of such systems is a multiple of a universal value $\sigma_{\textrm{min}}=e^2/(\pi h)$ independent of the material characteristics, if definite conditions are fulfilled and the chemical potential is tuned to the Dirac point~\cite{Katsnelson2006}. A minimal conductivity of the TI should be one quarter of those in graphene and is equal to $\sigma_{\textrm{min}}$ since there is no spin and valley degeneracy~\cite{Culcer2010}. However, experiments show that the minimal conductivity of TIs is larger than $\sigma_{\textrm{min}}$~\cite{Kim2012,Xu2016,Sacepe2011,Checkelsky2011}. It was attributed to the charge fluctuations near the Dirac point~\cite{Culcer2010} but recently in Ref.~\onlinecite{Peng2016}, it is argued that these fluctuations cannot explain properly the observed value of the conductivity in the TIs. Moreover, in the experiment~\cite{Banerjee2014} it has been observed unexpectedly that after creating more surface defects the conductivity in the (TI) exfoliated BiSbTeSe$_2$ nanoflakes increases~\cite{Banerjee2014}.

Magnetic field brings many interesting effects in the physics of TIs, including topological magnetoelectric effect~\cite{Qi2008}, existence of magnetic monopoles~\cite{Qi2009}, etc. In particular, the anomalous anisotropic magnetoresistance (AMR), that is, a dependence of the conductivity on the angle between applied magnetic field and current~\cite{McGuireJul1975}, is large in the TIs~\cite{Kandala2015,Pan2016,Taskin2017}. The value of the observed AMR cannot be explained by the common formula for the ferromagnetics. Note that the AMR in the TIs is large in the transverse (or out-of-plane) applied magnetic field as well as in the in-plane field~\cite{Kandala2015}. Another interesting phenomena occurred in the TIs is a quantum anomalous Hall effect (QAHE)~\cite{Liu2016}. When the out-of-plane magnetization is large, the TI exhibits a phase with the non-trivial first Chern number~\cite{Wang2015}. In this phase, a robust edge conductivity persists even if there is no flux quanta piercing the sample and the non-diagonal conductivity has a universal value $\sigma_0=e^2/(2h)$. The QAHE has been measured in thin films of the TIs and its value is equal to the theoretically predicted one with a good accuracy~\cite{Chang2013,Chang2015,Qi2016}. Recently, in Ref.~\onlinecite{Taskin2017} it has been reported the observation of the planar Hall effect in the TI, that is, Hall effect in the applied in-plane magnetic field.

\begin{figure}[t!]
\center
\includegraphics [width=8.5cm, height=4.5cm]{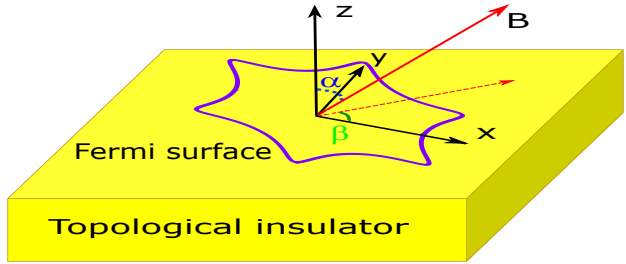}
\caption{Topological insulator with the hexagonal warping in the applied magnetic field. Violet line shows the Fermi surface, $x$ axis is directed along the scion of the Fermi surface, while $y$ axis is directed along the hollow of the surface.}
\label{scheme}
\end{figure}

The Dirac equation describes well the electronic states in the TI near the Dirac point. At higher energies, the Dirac cone transforms to a hexagonal snowflake, Fig.~\ref{scheme}. This effect is usually refereed to as a hexagonal warping. The value of this warping is controlled by the crystal structure of the TIs. The hexagonal warping is significant in Bi$_2$Te$_3$~\cite{Fu2009}. It is smaller in Bi$_2$Se$_3$~\cite{Kuroda2010}. Besides hexagonal warping, a massive quadratic term in the kinetic energy is also experimentally measured in the TIs. It destroys an electron-hole symmetry of the system. However, the effect of the latter term on the charge transport is not too large~\cite{Nomura2014}. The effect of the weak hexagonal warping on the transport properties of the TIs has been studied in Ref.~\onlinecite{Wang2011} and in Ref.~\onlinecite{Pal2012} with taking into account an electron-electron interaction. The influence of the hexagonal warping on the magnetic ordering of TI was analyzed in Ref.~\onlinecite{Siu2014}. The Landau levels in the TI with the hexagonal warping were calculated in Ref.~\onlinecite{Repin2015}.

Here we study the charge conductivity of the surface states in the TI with taking into account the hexagonal warping. We analyze the effects of disorder and magnetic field. We show that the existence of the hexagonal warping affects dramatically the appearance of the QAHE and AMR in the TI giving rise to a characteristic dependence of the off-diagonal and longitudinal conductivities on the direction and value of the magnetic field. We also demonstrate that the longitudinal conductivity increases with the increase of disorder in the TI with the significant hexagonal warping. The obtained results may be of importance for explanation of recently performed experiments.

The paper is organized as follows. In Section~\ref{model} we formulate the model, which will be used for conductivity computation. In Section~\ref{QAHE} we analyze the behavior of the off-diagonal conductivity and, particularly, the QAHE in the TI. In Section~\ref{longitud} we calculate the longitudinal conductivity and study the in-plane and out-of-plane AMR. In Section~\ref{discuss} we discuss the obtained results and, when possible, compare them with the experiment.

\section{The model}\label{model}

We consider the surface states in the TI placed in the Zeeman magnetic field. We  take into account the hexagonal warping and neglect Landau quantization. In this case the Hamiltonian of  the system is given by~\cite{Fu2009} ($\hbar=1$)
\begin{eqnarray}\label{H0}
\hat{H}=\mu\!+\!v_F(k_x \sigma_y\! -\! k_y \sigma_x)\! + \! \lambda k_x (k_x^2\!-\!3k_y^2)\sigma_z\!+\!{\bf B}\cdot{\bf \sigma},
\end{eqnarray}
where $\mathbf{\sigma}=(\sigma_x,\sigma_y,\sigma_z)$ are the Pauli matrices acting in the spin space, $\mu$ is the shift of the chemical potential from the Dirac point (which can be tuned either by gate voltage or doping), $v_F$ is the Fermi velocity, $k_x=k\cos \phi$ and $k_y=k\sin \phi$ are the in-plane momentum components, $\lambda$ is the strength of the hexagonal warping, ${\bf B}=(B_x,B_y,B_z)=(B_{\textrm{in}} \cos \beta,B_{\textrm{in}}\sin \beta,B_z)$ is the Zeeman magnetic field, $\alpha$ is the the azimuthal angle between the magnetic field and $z$ axis, $\beta$ is the angle of the in-plane component of the magnetic field $B_{\textrm{in}}$ with the crystal axis $x$ (see Fig.~\ref{scheme}). The term in the Hamiltonian responsible for the hexagonal warping can be rewritten as $\lambda k_x (k_x^2-3k_y^2)=\lambda k^3 \cos 3\phi$ and the Hamiltonian is invariant under rotation on the angle $\phi=2\pi/3$. The spectrum of the system is given by
\begin{eqnarray}\label{spectrum}
&&\!\!E_{\pm}=\mu \pm \epsilon\\
\nonumber
&&\!\!\epsilon=\sqrt{v_F^2k^2\!+\!B_{\textrm{in}}^2\!-\!2v_FkB_{\textrm{in}}\sin(\phi\!-\!\beta)\!+\!(\lambda k^3\cos 3\phi\!+\!B_z)^2}.
\end{eqnarray}
We can introduce a dimensionless parameter $\bar{\lambda}=\lambda \mu^2/v_F^3$, which is a measure of the strength of the hexagonal warping. If $\bar{\lambda}\ll1$, then, the linear term in the kinetic energy dominates over the hexagonal warping, while in the opposite case, $\bar{\lambda}\gg1$, the warping term is dominant.

The velocity operators takes the form
\begin{eqnarray}\label{velocity}
v_x=\frac{\partial \hat{H}}{\partial k_x}=v_F\sigma_y+3\lambda k^2\cos 2\phi\,\sigma_z, \nonumber\\
v_y=\frac{\partial \hat{H}}{\partial k_y}=-v_F\sigma_x-3\lambda k^2\sin 2\phi\,\sigma_z.
\end{eqnarray}
These operators are invariant under rotation on the angle $\phi=\pi$, which is different from the rotation symmetry of the Hamiltonian.

\section{Off-diagonal conductivity and QAHE}\label{QAHE}

At zero temperature and in the clean limit the off-diagonal conductivity $\sigma_{xy}$ can be written in terms of the Berry phase~\cite{Jungwirth2002,Haldane2004}
\begin{eqnarray}\label{sxy}
\sigma_{xy}=2\pi\sigma_0\int\limits_{\textrm{filled states}} d^2\,k\,(\partial_x A_y^m -\partial_y A_x^m),
\end{eqnarray}
where $A_l^m=i\langle u_m|\partial_l|u_m\rangle$ is the Berry connection and $u_m$ is the Bloch vector of the band $m$. If the spectrum is gapped, then, the integration in Eq.~\eqref{sxy} is taken over the whole Brillouin zone, and the off-diagonal conductivity is quantized, that is,  $\sigma_{xy}=C_1\sigma_0$, where $C_1$ is the first Chern number, that is, integer.

\begin{figure}[t!]
\center
\includegraphics [width=8.5cm]{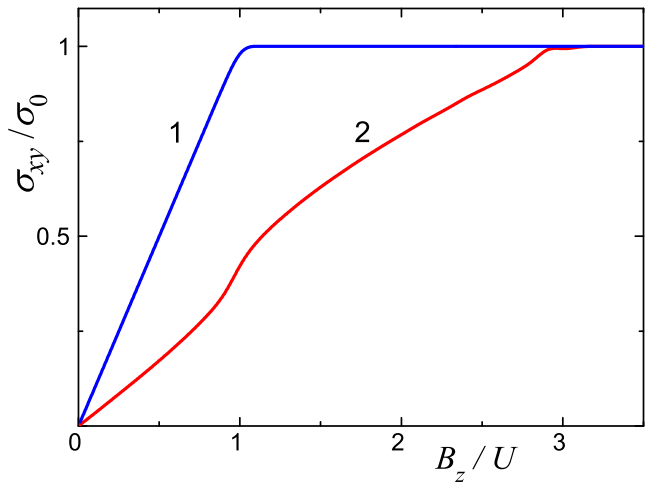}
\caption{Off-diagonal conductivity $\sigma_{xy}$ in the out-of-plane magnetic field $B_z$ for $\lambda=0$ (blue) line 1 and $\lambda \mu^2/v_F^3=3$ (red) line 2.}
\label{sxyvz}
\end{figure}

\begin{figure}[t!]
\center
\includegraphics [width=8.5cm]{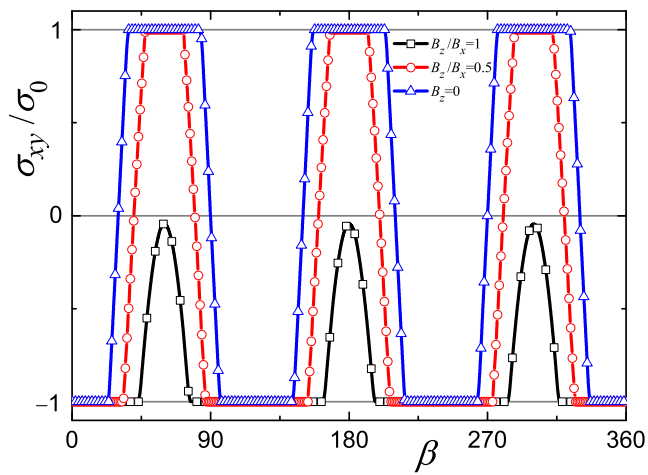}
\caption{The dependence of the off-diagonal conductivity $\sigma_{xy}$ on the angle $\beta$ between the in-plane magnetic field and the crystal axis $x$ for $\lambda \mu^2/v_F^3=0.1$, $B_{\textrm{in}}/\mu=10$, and different values of the out-of-plane field $B_z$: $B_z/B_{\textrm{in}}=0$ corresponds to (blue) line with up triangles, $B_z/B_{\textrm{in}}=0.5$ corresponds to (red) line with circles, and $B_z/B_{\textrm{in}}=1$ corresponds to (black) line with squares.}
\label{stm_exp}
\end{figure}

\begin{figure}[t!]
\center
\includegraphics [width=8.5cm]{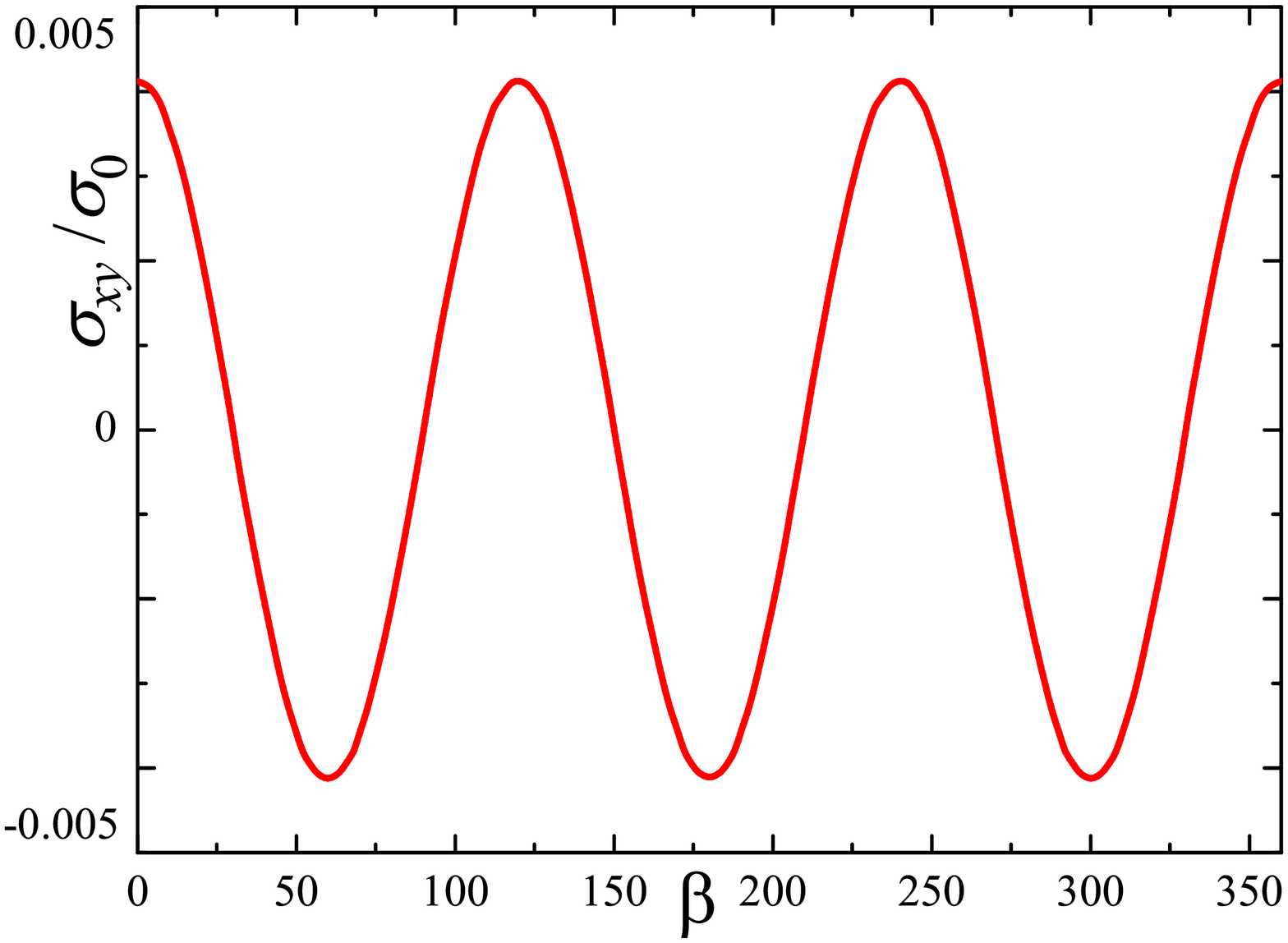}
\caption{The dependence of the off-diagonal conductivity $\sigma_{xy}$ on the angle $\beta$ between the in-plane magnetic field and the crystal axis $x$ in the case of large chemical potential $\mu$; $\lambda \mu^2/v_F^3 = 4$ and $B_{\textrm{in}}/\mu=0.5$.}
\label{phe}
\end{figure}

According to Eq.~\eqref{spectrum}, the application of large transverse magnetic field $B_z$ opens a gap in the spectrum of the surface electrons and the QAHE will be observed. In the case of small hexagonal warping, that is, $\bar{\lambda}\ll 1$, the (gapped) topologically nontrivial phase with $|C_1|=1$ arises if $B_z>\mu$. In the case, when the hexagonal warping is large, $\bar{\lambda} \gg 1$, the gap opens if $B_z \gtrsim\bar{\lambda} \mu$. Therefore, the magnetic field at which the off-diagonal conductivity is quantized increases with the increase of the hexagonal warping. The sign of the quantized plateaus in $\sigma_{xy}$ changes if we change the sign of $B_z$. The dependence of the conductivity $\sigma_{xy}$ on the transverse magnetic is shown in Fig.~\ref{sxyvz} for two different values of  $\bar{\lambda}$. Note, we assume that the Hamiltonian~\eqref{H0} is valid over the whole Brillouin zone when calculating curves in Fig.~\ref{sxyvz}.

Application of the in-plane magnetic field $B_\textrm{in}$ gives rise only to the shift of the Dirac cone in the $k$-space and does not effect the value of $\sigma_{xy}$ if we neglect the hexagonal warping. However, when we take into account that $\lambda\neq0$, we readily reveal that the spectrum becomes gapped if the in-plane magnetic field is sufficiently large ($\lambda B^2_{\textrm{in}}/v_F^3\sim 1$), as it follows from Eq.~\ref{spectrum}. We calculate the conductivity $\sigma_{xy}$ as a function of the angle $\beta$ between the in-plane field and the crystal axis $x$ by means of Eqs.~\eqref{spectrum} and \eqref{sxy} for different values of parameters. The results are illustrated in Figs.~\ref{stm_exp}, where a rather peculiar picture of the QAHE is observed. In the case $B_z=0$ (blue line with up-triangles) the symmetry of the energy spectrum corresponds to a rotation on the angle $\pi/3$. Therefore, the conductivity $\sigma_{xy}(\beta)$ has the same symmetry. If the in-plane field $B_{\textrm{in}}$ is large, the off-diagonal conductivity achieves quantized plateaus $\sigma_{xy}=\pm \sigma_0$ in definite range of angles, where the gap arises in the electron spectrum. The range of these angles increases with the growth of $B_{\textrm{in}}$. Application of the out-of-plane field $B_z$ reduces the symmetry of the Fermi surface from $\pi/3$ to $2\pi/3$ and, consequently, breaks the symmetry of $\sigma_{xy}$ with respect to $C_1=1$ and $C_1=-1$. In Fig.~\ref{stm_exp}, we plot the conductivity $\sigma_{xy}$ as a function of $\beta$ in the case $B_z>0$. The range of angles $\beta$, where $\sigma_{xy}=\pm\sigma_0$ decreases with the growth of $B_z$. The quantized plateau in $\sigma_{xy}$ disappears if $B_z$ is significantly large, the curve with $B_z/B_{\textrm{in}}=1$ in Fig.~\ref{stm_exp}. 

The off-diagonal conductivity decays with the increase of the chemical potential $\mu$. If $\mu$ is large, the off-diagonal conductivity is strongly suppressed and the angle dependence of $\sigma_{xy}$ can be fitted by $\cos {3\beta}$ rule, see Fig.~\ref{phe}.

Let us emphasize once more, that any dependence of $\sigma_{xy}$ on the angle $\beta$ disappears if we neglect the hexagonal warping.

\section{Longitudinal conductivity}\label{longitud}

In this section we calculate the longitudinal conductivity $\sigma_{ii}$ ($i=x,y$) of the TI in the magnetic field. For this goal, we apply Bastin-Kubo-Streda formula in the form~\cite{Proskurin2015}
\begin{eqnarray}\label{sxx}
\sigma_{\alpha \alpha}=2\pi\sigma_0\sum_\mathbf{k}\textrm{Tr} [V_{\alpha}\, \textrm{Im} G \,v_{\alpha}  \, \textrm{Im} G],
\end{eqnarray}
where $V_{\alpha}$ is the vertex corrected velocity operator, which is discussed in Appendix, $\textrm{Im} G =  i(G^{+}-G^{-})/2$ and $G^{\pm}$ are the impurity averaged Green's functions. These functions are calculated in Appendix. They are given by Eq.~\eqref{greenF} in the case of the point-like random disorder and not too high Zeeman magnetic field, $B<\mu$. Since the vertex corrections vanishes for the short range disorder~\cite{Bastin1971,Shon1998}, all information about disorder is given by scattering amplitude $\Gamma$ in the impurity averaged Green functions, Eq.~\eqref{greenF}. Note, the longitudinal conductivity is isotropic if $B_{\textrm{in}}=0$, as it follows from Eqs.~\eqref{sxx} and~\eqref{greenF}.

We start with the simplest case of zero magnetic field and $\mu=0$ (the Fermi level passes through the Dirac point). If we neglect the hexagonal warping, the conductivity takes a universal value $\sigma_{xx}=\sigma_{yy}=\sigma_0$, which is quarter of the minimal conductivity of graphene and is independent of the disorder~\cite{Culcer2010}. If we take into account the hexagonal warping, the ratio $\sigma_{ii}/\sigma_0$ depends on a single parameter $\lambda \Gamma^2/v_F^3$, as it follows from Eqs.~\eqref{velocity}, \eqref{sxx}, and \eqref{greenF}. Calculations show that the conductivity increases monotonically with the increase of the value $\lambda \Gamma^2/v_F^3$, that is, with the grows of the both hexagonal warping and disorder. This dependence can be approximated as $\sigma_{xx}=\sigma_{yy}\approx \sigma_{\textrm{min}}(1+13\lambda \Gamma^2/v_F^3)$ when $\lambda \Gamma^2/v_F^3<1$.

\begin{figure}[t!]
\center
\includegraphics [width=8.5cm]{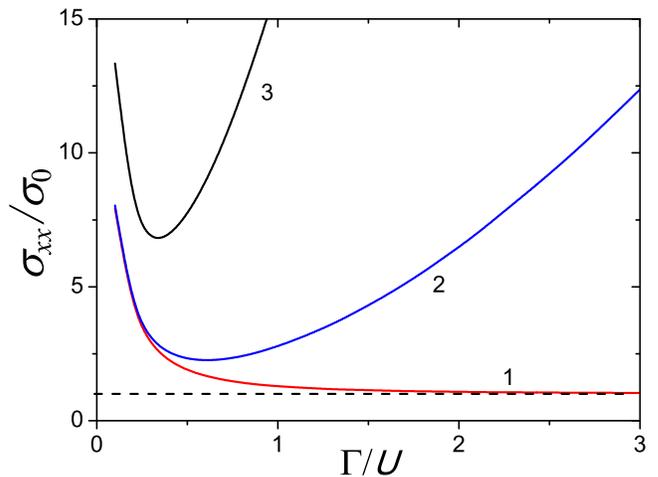}
\caption{The dependence of the longitudinal conductivity $\sigma_{xx}$ on the disorder parameter $\Gamma/\mu$ in the case of $\mathbf{B}=0$ for different values of the hexagonal warping; (red) line 1 corresponds to $\lambda=0$, (blue) line 2 to $\lambda \mu^2/v_F^3=0.1$, and (black) line 3 to $\lambda \mu^2/v_F^3=1$. Dash line indicates the minimum conductivity $\sigma_{\textrm{min}}$.}
\label{sg}
\end{figure}

If the system is shifted from the charge neutrality point, $\mu\neq0$, the conductivity $\sigma_{ii}$ depends on two dimensionless parameters. It is convenient to choose them as $\Gamma/\mu$, which characterizes disorder, and $\lambda\mu^2/v_F^3$, which characterizes the hexagonal warping. The results of the calculations of $\sigma_{xx}$ as a function of $\Gamma/\mu$ at different $\lambda\mu^2/v_F^3$ with the help of Eqs.~\eqref{velocity}, \eqref{sxx}, and \eqref{greenF} are shown in Fig.~\ref{sg}. As it is seen from the figure, the conductivity $\sigma_{ii}$ decreases with disorder at low $\Gamma/\mu$. The behavior of $\sigma_{ii}$ at larger $\Gamma/\mu$ depends on the hexagonal warping. If the warping is negligibly small, the increase of the disorder, $\Gamma\rightarrow\infty$, is equivalent to the shift of the system toward the charge neutrality point, $\mu \rightarrow 0$. Thus, the increase of the impurities concentration reduces the conductivity down to the minimal value $\sigma_0$, when $\Gamma/\mu\gg 1$ (see curve 1 in Fig.~\ref{sg}). A presence of the hexagonal warping changes the behavior of the conductivity at large disorder drastically (curve 2 and 3 in Fig.~\ref{sg}). The function $\sigma_{xx}(\Gamma/\mu)$ drops to minimum at $\Gamma/\mu\sim 1/2$ and then increases with the increase of the disorder.

\begin{figure}[t!]
\center
\includegraphics [width=8.5cm]{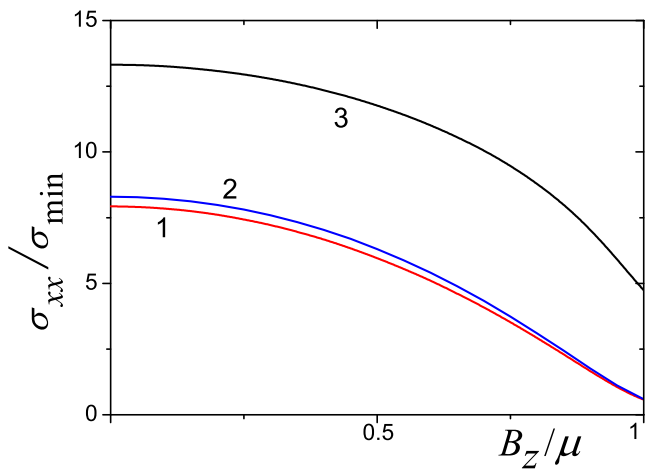}
\caption{The dependence of the longitudinal conductivity $\sigma_{xx}$ on the out-of-plane magnetic field $B_z$ for $\Gamma/\mu=0.1$ and different values of the hexagonal warping;  (red) line 1 corresponds to $\lambda=0$, (blue) line 2 to $\lambda \mu^2/v_F^3=0.1$, and (black) line 3 to $\lambda \mu^2/v_F^3=1$.}
\label{sbz}
\end{figure}

The application of the out-of-plane magnetic field $B_z$ reduces the density of states at the Fermi level, which leads to the reduction of the longitudinal conductivity. The conductivity $\sigma_{xx}$ as a function of the field $B_z$ is shown in Fig.~\ref{sbz} for different values of the hexagonal warping.

\begin{figure}[t!]
\center
\includegraphics [width=8.5cm]{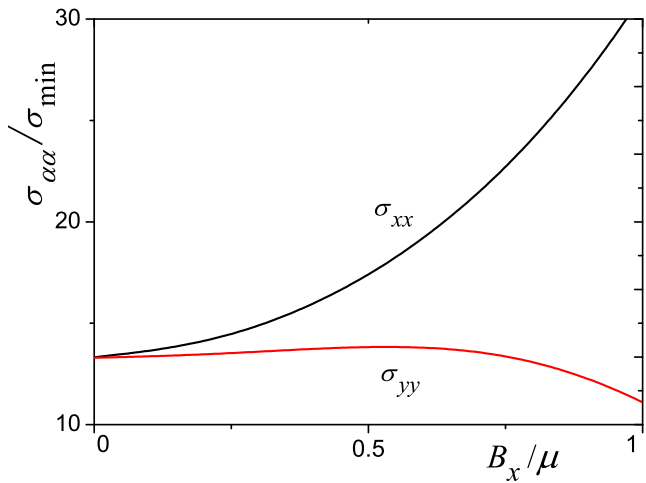}
\caption{The components of the longitudinal conductivity $\sigma_{xx}$ (upper black curve) and $\sigma_{yy}$ (lower red curve) shown as functions of the in-plane magnetic field $B_x$; $\lambda \mu^2/v_F^3=1$ and $\Gamma/\mu=0.1$.}
\label{sas}
\end{figure}

As it was mentioned in the previous section, the in-plane magnetic field only shifts the Dirac point in the $(k_x,k_y)$ plane and does not effect the transport properties if the hexagonal warping is neglected. If we take into account the hexagonal warping, we observe a significant in-plane AMR, that is, the conductivity becomes anisotropic and magnetic field dependent. The calculated components of the longitudinal conductivity $\sigma_{xx}$ and $\sigma_{yy}$ as functions of $B_x$ are shown in Fig.~\ref{sas}. As it follows from the figure, the conductivity in the direction of the magnetic field increases with the increase of the in-plane field, while the conductivity in the perpendicular direction decreases.

\begin{figure}[t!]
\center
\includegraphics [width=8cm]{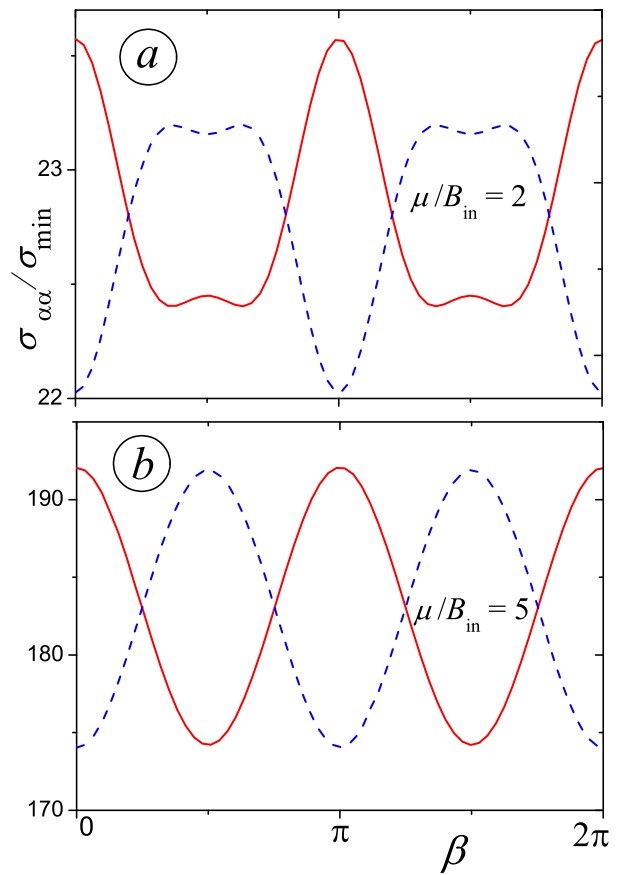}
\caption{The longitudinal conductivities $\sigma_{xx}$ [(red) solid lines] and $\sigma_{yy}$ [(blue) dash lines] as functions of the angle $\beta$ between the in-plane magnetic field $\mathbf{B}_{\textrm{in}}$ and the crystal axis $x$ at different $\mu$: $\mu/B_{\textrm{in}}=2$ in panel $a$ and $\mu/B_{\textrm{in}}=5$ in panel $b$ ($B_z=0$, $\Gamma/B_{\textrm{in}}=0.1$, $\lambda B_{\textrm{in}}^2/v_F^3=0.15$). In the absence of the hexagonal warping ($\lambda=0$), the anisotropy disappears and $\sigma_{xx}=\sigma_{yy}$;  $\sigma_{xx}/\sigma_{\textrm{min}}=15.746$ in the case $\mu/B_{\textrm{in}}=2$ and $\sigma_{xx}/\sigma_{\textrm{min}}=39.25$ in the case $\mu/B_{\textrm{in}}=5$.}
\label{sl1}
\end{figure}

In Fig.~\ref{sl1} we plot the dependence of the longitudinal conductivity components $\sigma_{xx}$ and $\sigma_{yy}$ on the angle $\beta$ between the direction of the in-plane magnetic field $\mathbf{B}_{\textrm{in}}$ and the crystal axis $x$. First, we have to emphasize that the anisotropy of $\sigma_{\alpha\alpha}(\beta)$ exists only due to the hexagonal warping. We see from Figs.~\ref{sl1} $a$ and $b$ that the observable peaks in the conductivities $\sigma_{xx}(\beta)$ and $\sigma_{yy}(\beta)$ arise at the angles $\beta =N\pi/3$ where $N$ is integer, if the system is not far from the charge neutrality point (panel $a$). When the Fermi level of the system is shifted far from the Dirac point, the  peaks in $\sigma_{xx}(\beta)$ arises if $\beta=\pi N$ and in $\sigma_{yy}(\beta)$ if $\beta=\pi (N+1/2)$ (panel $b$). Varying $\mu$ by the gate voltage, we can change the symmetry of the conductivity from nearly rotation on $\pi/3$ to nearly $\pi$. This effect occurs due to the difference in the rotational symmetries of the model Hamiltonian Eq.~\eqref{H0} and the velocity operators Eq.~\eqref{velocity}. Note also that the conductivity increases with the increase of $\mu$ but the relative anisotropy of the conductivity and, consequently the in-plane AMR, decreases in this case. Note also that the AMR evidently increases with the growth of $B_{\textrm{in}}$.

\begin{figure}[t!]
\center
\includegraphics [width=8.5cm]{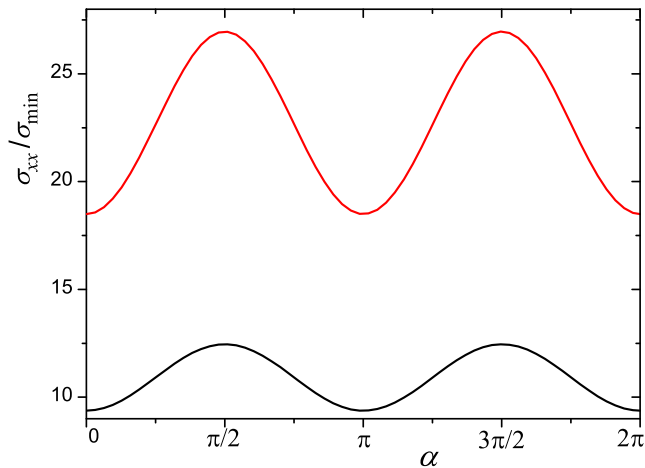}
\caption{The dependence of the longitudinal conductivity $\sigma_{xx}$ on the azimuthal angle $\alpha$ between $z$ axis and the applied magnetic field $\mathbf{B}=(B\sin \alpha,0,B\cos \alpha)$. Lower (black) line corresponds to $\lambda=0$ (absence of the hexagonal warping) and in the upper (red) line $\lambda B^2/v_F^3=1$; $\Gamma/\mu=0.1$ and $B/\mu=1$.}
\label{salpha}
\end{figure}

The diagonal conductivity $\sigma_{ii}$ also depends on the azimuth angle $\alpha$ between $z$ axis and direction of the magnetic field (the out-of-plane AMR). This is illustrated in Fig.~\ref{salpha}, where the dependence of $\sigma_{xx}$ on $\alpha$ is shown at different values of the hexagonal warping. We obtain that the conductivity approximately follows a typical $\sin^2 \alpha$ rule. The value of the conductivity increases with the increase of the warping.

\section{Discussion}\label{discuss}

In this paper we study the effect of the hexagonal warping on the conductivity of the three-dimensional TI. We observe that this effect is significant, especially, when the magnetic field is applied. However, in our calculations we used specific values of the system parameters and it is an important question, whether these values are realistic for existing samples. For the estimates we base on the data on the TI electronic spectra from Ref.~\onlinecite{Nomura2014} and a characteristic value of the disorder $\Gamma = 22$~meV from Ref.~\onlinecite{Beidenkopf2011} (see also speculations below Table~\ref{my-label}).

In our consideration we neglect the orbital effect of the magnetic field. This is evidently justified when we analyze the system in the in-plane field. Naturally, for the in-plane magnetic field we can choose the gauge for the vector potential $\mathbf{A}=(0,0,A_z)$. Since $k_z$ is absent in the Hamiltonian, then, the Pierls substitution $k_z\rightarrow k_z -e/c A_z$ does not change anything. The orbital effect of the out-of-plane magnetic field leads to the Landau quantization. Thus, when the out-of-plane field is taken into consideration, we must restrict our treatment by the case, when $B_z$ is not too high and the Landau quantization is absent. In the experiments we discuss below, the Shubnikov-de Haas oscillations have not been observed, so we can safely neglect orbital effects of the magnetic field. Note also that the Landau levels can be smeared out in a disordered system~\cite{Cheng2010}.

In the Hamiltonian~\eqref{H0} the chemical potential $\mu$ is assumed to be zero when the Fermi level passes through the Dirac point. In real samples, the value of $\mu$ depends on the sample composition or doping. In experiments, this value can be tuned by application of a gate voltage $V_g$. At some $V_g=V_g^0$ the chemical potential $\mu=0$ and the sample is at the charge neutrality position~\cite{Kandala2015}. Thus, we can write $\mu=V_g-V_g^0$ and use the measured value of $V_g^0$ as a reference value of $\mu$ for the estimate of the hexagonal warping parameter $\lambda \mu^2/v_f^3$ and disorder parameter $\Gamma/\mu$. In Table~\ref{my-label} we present the values of $\lambda \mu^2/v_f^3$ and $\Gamma/\mu$ extracted from the experimental date presented in Ref.~\onlinecite{Nomura2014} for two different TIs. Within the order of magnitude, these parameters are close to that used us in the previous sections for estimates.

\begin{table}[]
\centering
\caption{The values of the parameters extracted from the experimental date in Ref.~\cite{Nomura2014}}
\label{my-label}
\begin{tabular}{|l|l|l|}
\hline
 & $\lambda \mu^2/v_F^3$
 & $\Gamma/\mu$
 \\ \hline
Bi$_2$Se$_3$ & 0.5 & $0.07$ \\ \hline
 Bi$_2$Te$_3$ &  0.35 &   $0.065$        \\ \hline
\end{tabular}
\end{table}

The estimated dimensionless parameters characterizing the warping and disorder are close for Bi$_2$Se$_3$ and Bi$_2$Te$_3$ samples, as it follows from the data presented in Table~\ref{my-label}. Thus, the effect of the hexagonal warping will be similar for these two compounds. Really, the observed values of the AMR for bismuth telluride~\cite{Kandala2015} and bismuth selenide ~\cite{Pan2016} are of the same order.

The model Hamiltonian~\eqref{H0} obeys an electron-hole symmetry. This symmetry will be broken if we add a quadratic term $r(k_x^2+k_y^2)$ in the Hamiltonian. For the parameter $r$ extracted from the ARPES data in Ref.~\onlinecite{Nomura2014} we found by direct computations that the quadratic correction does not affect significantly the results obtained above except a slight violation of the electron-hole symmetry, that is, the symmetry with respect to sign of $\mu$.

Making sure that the values of parameters used above are reasonable, we continue to discuss the obtained results.

The hexagonal warping affects dramatically the QAHE in the TI, that is, the behavior of the off-diagonal conductivity in the magnetic field (Section~\ref{QAHE}). In particular, the oscillations of the quantized conductivity $\sigma_{xy}$ with the rotation of the in-plane magnetic field are possible only due to the hexagonal warping (see Fig.~\ref{stm_exp}). If one neglects the hexagonal warping the effect of the applied in-plane magnetic field reduces to the shift of the Dirac cone in the $k$-space. The out-of-plane component of the magnetic field $B_z$ opens the gap in the electron spectrum and gives rise to the quantization of the off-diagonal conductivity whether the hexagonal warping is of significance or not. However, the gaps in the spectrum and quantization of $\sigma_{xy}$ will be observed in higher field in the samples with larger warping (Fig.~\ref{sxyvz}). Moreover, we can change the appearance of the QAHE in the in-plane field varying the value and direction of $B_z$ (cf. curves for $B_z=0$ and $B_z>0$ in Fig.~\ref{stm_exp}). If the shift of the system from the Dirac point is significant (that is, $\mu$ is large) or the in-plane magnetic field is small, the off-diagonal conductivity is suppressed in comparison with QAHE regime. Yet, finite off-diagonal conductivity can be measured and the angle dependence of $\sigma_{xy}$ follows $\cos{(3\beta)}$ rule. Note that in the recent experiment in Ref.~\onlinecite{Taskin2017} the planar Hall effect in TI was observed. However, in this experiment the transverse conductivity follows the $\cos{(2\beta)}$ rule which is different from our prediction. Thus, the problem needs further investigation.

The hexagonal warping may be responsible for several important features in the dependence of the longitudinal conductivity on disorder and applied magnetic field (Section~\ref{longitud}). The presence of the hexagonal warping gives rise to a nonmonotonic dependence of the longitudinal conductivity on the disorder strength $\Gamma$ and, in particular, the growth of $\sigma_{xx}$ with disorder at large $\Gamma$, see Fig.~\ref{sg}. This result can explain why certain topological insulators have minimal conductivity larger than expected~\cite{Sacepe2011,Kim2012}. Moreover, a recent experiment shows that the conductivity of TI really growth with the increase of disorder\cite{Banerjee2014}. At present, we cannot find a clear physical explanation of this effect. We performed quasi-classical calculations and obtain that the diffusion coefficient of the system is only slightly renormalized by the hexagonal warping. A similar result have been obtained in Ref.~\onlinecite{Adroguer2012}. So, we conclude that the effect is purely quantum. Mathematically, we check that the growth of the longitudinal conductivity with disorder is a consequence of the term $\propto k_{\alpha}^2 \sigma_z$ in Eq.~\eqref{velocity} for the velocity operator $v_{\alpha}$. In fact, the anomalous growth of the conductivity appears in a more general case, say, if $v_{\alpha} \propto k_{\alpha}^n \sigma_z$ where $n>0$. This issue deserves a future study. Note, that an increase of the conductivity with the increase of disorder was reported in Ref.~\onlinecite{Peng2016}, where the conductivity was calculated for a thin film of TI using a tight binding model. However, in that case the growth of $\sigma_{xx}$ with the disorder occurs due to interaction of the electrons in different layers of the film.

In fact, Dirac systems are known to be insensitive to the strong localization (if electron-electron or electron-phonon coupling are not strong~\cite{Bardarson2007}). Otherwise, the system with Dirac dispersion would be Anderson insulator at the Dirac point. The Klein tunneling and absence of the backscattering are the reasons for the suppression of localization in the Dirac systems. So, our calculations are applicable if disorder does not mix the Dirac and bulk states in the TI and the obtained results are valid even when $\Gamma>\mu$.

The hexagonal warping gives rise to the anisotropy of the longitudinal conductivity in the applied in-plane magnetic field, see Fig.~\ref{sas}. The ratio of the conductivities along and transverse the direction of in-plane field increases with the increase of $B_{\textrm{in}}$. The hexagonal warping is also responsible for the AMR in the in-plane field, see Figs.~\ref{sl1}. The dependence of $\sigma_{xx}$ and $\sigma_{yy}$ on the direction of the applied in-plane field has an oscillating nature reflected the mixture of rotational symmetries of the Hamiltonian, Eq.~\eqref{H0}, ($\pi/3$) and electron velocity, Eq.~\eqref{velocity}, ($\pi$). The appearance of these oscillations strongly depends on the chemical potential and the value of the warping. If the the system is shifted from the Dirac point ($\mu\neq 0$) the in-plane AMR can be approximated by a typical $\cos^2\beta$ law (see curves in Fig.~\ref{sl1}$b$), as it has been observed experimentally~\cite{Kandala2015,Pan2016,Taskin2017}. However, in general, $\beta$-dependence of the AMR may be far from $\cos^2\beta$ (Fig.~\ref{sl1}$a$).

The longitudinal conductivity, as well as off-diagonal one, is sensible to the value of the out-of-plane component $B_z$ of the magnetic field. Due to a specific spin structure of the Dirac Hamiltonian, the growth of $B_z$ reduces the density of states in the Fermi level and, consequently, gives rise to the decrease of the longitudinal conductivity, see Fig.~\ref{sbz}. As a result, a sharp peaks are observed in the longitudinal conductivity when the angle $\alpha$ between $z$ axis and the applied magnetic field $\mathbf{B}$ is equal to $\alpha=\pi/2$ or $3\pi/2$ (see Fig.~\ref{salpha}), that is, when $B_z=0$. The value of the hexagonal warping affects the value of the effect.

\begin{figure}[t!]
\center
\includegraphics [width=8.5cm]{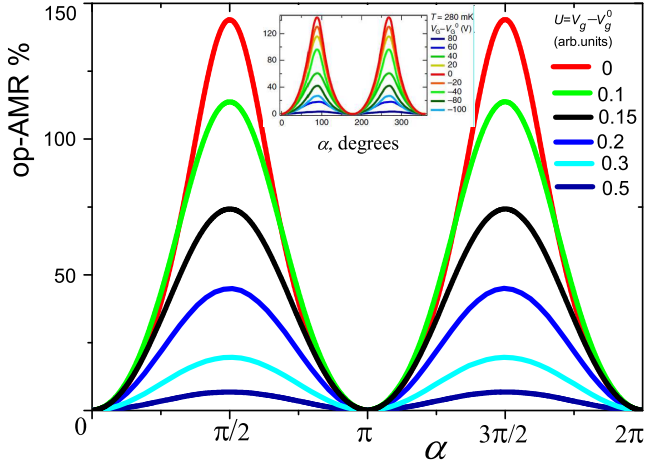}
\caption{The dependence of the anomalous out-of-plane magnetoresistance,  $\textrm{op-AMR}=(\sigma_{xx}^{\textrm{max}}-\sigma_{xx}^{\textrm{min}})/\sigma_{xx}^{\textrm{min}}$), on the angle $\alpha$ between $z$ axis and the applied magnetic field; $\lambda \mu^2/v_F^3=0.5$, $\Gamma/\mu=0.1$, and $B/\mu=0.13$. The experimental data from Ref.~\onlinecite{Kandala2015} (Fig.4$a$ from this reference) are shown in the inset.}
\label{Gate_alpha}
\end{figure}

\begin{figure}[t!]
\center
\includegraphics [width=8.5cm]{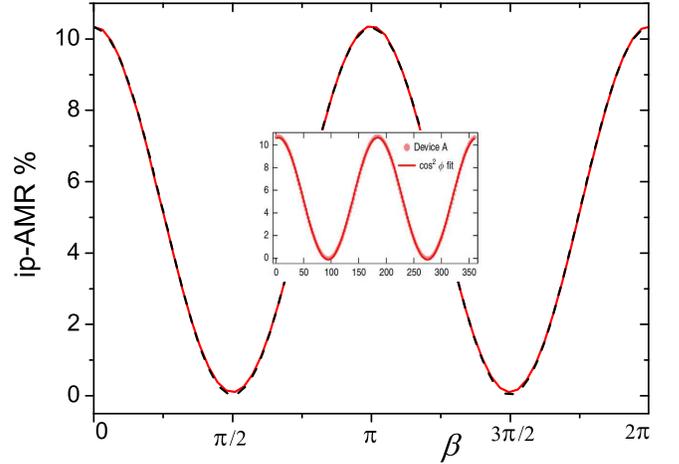}
\caption{Solid (red) curve shows the dependence of the anomalous in-plane magnetoresistance,  $\textrm{in-AMR}=(\sigma_{xx}^{\textrm{max}}-\sigma_{xx}^{\textrm{min}})/\sigma_{xx}^{\textrm{min}}$), on the angle $\beta$ between $x$ axis and the in-plane magnetic field $\mathbf{B}_{\textrm{in}}$. The result is extracted from the data shown in Fig.~\ref{sl1} $b$ ($\Gamma/B_{\textrm{in}}=0.1$, $\lambda B_{\textrm{in}}^2/v_F^3=0.15$, and $\mu/B_{\textrm{in}}=5$). The fit by $\cos^2{\beta}$ is shown by (black) dash curve. The experimental data from Ref.~\onlinecite{Kandala2015} (Fig.2$e$ from this reference) are shown in the inset.}
\label{inAMR}
\end{figure}

Anomalous out-of-plane AMR (op-AMR) and in-plane AMR (in-AMR) were observed in epitaxial films of Cr$_x$(Bi,Sb)$_{2-x}$Te$_3$ at low temperatures~\cite{Kandala2015}. The values of the op-AMR and ip-AMR  are defined here as $(\sigma_{xx}^{\textrm{max}}-\sigma_{xx}^{\textrm{min}})/\sigma_{xx}^{\textrm{min}}$. The dependence of op-AMR on the angle $\alpha$ calculated at plausible values of parameters is shown in Fig.~\ref{Gate_alpha} for different values of the chemical potential $\mu$. The calculated curve in-AMR versus $\beta$ is presented in Fig.~\ref{inAMR}. We show the experimental data obtained in Ref.~\onlinecite{Kandala2015} as the insets in these figures. A similarity between the theoretical and experimental results is very reasonable. Note, that in the experiment the measured quantity is the resistivity $\rho_{xx}=\sigma_{xx}/(\sigma_{xx}^2+\sigma_{xy}^2)$, which is inverse proportional to the longitudinal conductivity $\rho_{xx} \simeq 1/\sigma_{xx}$ if $\sigma_{xx} \gg \sigma_{xy}$. The latter condition is satisfied if $\mu > B$ and $\mu \gg \Gamma $, which is true for the typical parameters of the samples.

In conclusion, the hexagonal warping can significantly affect the transport properties of the topological insulators. In particular, it should be taken into account when considering quantum anomalous Hall effect, in-plane and out-of-plane magnetoresistance and effects of disorder. This issue can be a key for understanding of recently obtained experimental results.

\section*{Appendix}

Here we derive expressions for the impurity averaged Green's functions $G^\pm$. We start with the case of zero magnetic field ${\bf B}=0$. We write down a scattering potential in the form
\begin{equation}
V_{imp}=u_0 \sum\limits_i \delta(r-R_i),
\end{equation}
where $R_j$ are the positions of the randomly distributed point-like impurities with the local potential $u_0$ and concentration $n_i$. We assume that the disorder is Gaussian $\langle V_{imp} \rangle=0$ and $\langle V_{imp}(r_1) V_{imp}(r_2) \rangle=n_i u_0^2 \delta (r_1-r_2)$. Here $\langle ... \rangle$ stands for the impurity average and $\delta(r)$ is the Dirac delta function.

In the self-consistent Born approximation (SCBA), the impurity-averaged Green's functions $G^{\pm}$ can be calculated as
\begin{eqnarray}\label{dyson}
G^{\pm}=G_0^{\pm}+G_0^{\pm}\Sigma^{\pm} G^{\pm}
\end{eqnarray}
where $G_0^{\pm}$ are bare Green's functions of the Hamiltonian~\eqref{H0}
\begin{widetext}
	\begin{eqnarray}\label{green0}
	G^{\pm}_0=\frac{\mu\pm i0\!-\!v_F(k_x \sigma_y - k_y \sigma_x)\! - \! \lambda k_x (k_x^2-3k_y^2)\sigma_z\!}{(\mu\pm i0)^2-\left\{(v_Fk_x)^2+(v_Fk_y)^2+\left[\lambda k_x(k_x^2-3k_y^2)\right]^2\right\}}
	\end{eqnarray}
\end{widetext}
and $\Sigma^{\pm}$ is the self-energy, which is defined as
\begin{eqnarray}
\Sigma^{\pm}=  \langle V_{imp} G^{\pm} V_{imp}\rangle.
\end{eqnarray}
In the case under consideration, we can calculate the self-energy $\Sigma^{\pm}=\Sigma'\mp i\Gamma$ using the expression derived in Ref.~\onlinecite{Shon1998}
\begin{eqnarray}\label{SelfEnergy}
\!\!\Sigma\!=\!\frac{n_i u_0^2}{(2\pi)^2} (\mu\!-\!\Sigma)\!\!\!\! \int\limits_0^{+\infty}\!\int\limits_0^{2\pi}\!\!\frac{k dk\,d\phi }{(\mu\!-\!\Sigma)^2\!-\!v_F^2k^2\!-\!\lambda^2k^6\cos^2{3\phi}}.
\end{eqnarray}
The self-energy is proportional to the identity matrix. Therefore, the expression for $G^\pm$ is given by an equation similar to Eq.~\eqref{green0} for $G^\pm_0$, in which $\pm i0$ is replaced by $\pm \Gamma$. In general case, we can not derive an explicit expression for $\Sigma^{\pm}$ and consider two limiting cases of large and zero chemical potential $\mu$.

In the limit $\mu \gg |\Sigma|$, we can put $\Sigma' =0$ and assume that the imaginary part of the self energy $\Gamma$ is small. In this case we get from Eq.~\eqref{SelfEnergy}
\begin{equation}\label{Gamma1}
\Gamma=\frac {n_i u_0^2 |\mu|}{2\pi v_F^2}\left\{
                                            \begin{array}{ll}
                                              \frac {1}{\sqrt{1+\lambda^2 \mu^4/v_F^6}}, & \hbox{$\frac{\lambda \mu^2}{v_F^3}\ll 1$;} \\
                                              \frac {4 \pi^{1/2}}{3\Gamma(\frac 56)\Gamma(\frac 53) (\lambda \mu^2/v_F^3)^{2/3}}, & \hbox{$\frac{\lambda \mu^2}{v_F^3}\gg 1$,}
                                            \end{array}
                                          \right.
\end{equation}
where $\Gamma(x)$ is a gamma function and should not be confused with the scattering amplitude $\Gamma$. Note, that $\Gamma$ decreases with increase of chemical potential if the hexagonal warping is strong.

At the Dirac point, $\mu=0$, we get that self-energy has only imaginary part and Eq.~\eqref{SelfEnergy} reads
\begin{eqnarray}\label{integral}
1=\frac{n_i u_0^2}{(2\pi)^2}  \int\limits_0^{+\infty}\int\limits_0^{2\pi}\frac{k dk\,d\phi }{\Gamma^2+v_F^2k^2+\lambda^2 k^6 \cos^2{3\phi}}.
\end{eqnarray}
If we neglect hexagonal warping, $\lambda=0$, the integral in Eq.~\eqref{integral} diverges and we must put a finite cut-off value $k_c$ in the upper limit in this integral. After that we get
\begin{eqnarray}
\Gamma=\frac{v_F k_c }{\sqrt{e^{v_F^2/(2\pi n_i u_0^2)}-1}}.
\end{eqnarray}
A similar result has been obtained for the graphene in Ref.~\onlinecite{Hu2008}. In the case of small but finite hexagonal warping, $\lambda \Gamma^2/v_F^3 \ll 1$, the integral converges and we derive
\begin{eqnarray}
\Gamma=\sqrt{\frac{v_F^3}{\lambda}} e^{-v_F^2/(4 \pi n_i u_0^2)}
\end{eqnarray}

In the case of strong hexagonal warping, $\lambda \Gamma^2/v_F^3 \gg 1$, we find that
\begin{eqnarray}
\Gamma = \sqrt{\frac{v_F^3}{\lambda}} \left[\frac{n_i u_0^2}{\pi v_F^2}\frac{4\pi^{3/2}}{27\Gamma(\frac{5}{6})\Gamma(\frac{4}{3})}\right]^{3/4}.
\end{eqnarray}

 Vertex corrections can be calculated as follows
 \begin{eqnarray}
 V_{\alpha}=v_{\alpha}+\frac{n_i u_0^2}{(2\pi)^2} \int G^+ V_{\alpha} G^{-}
 \end{eqnarray}
 We found that $V_{\alpha}=2v_{\alpha}$, as it have been obtained in Refs, and hexagonal warping has a little effect on renormalized velocity operator in case of $U \gg \Gamma(\mu=0)$.
  
Thus, we related the scattering amplitude $\Gamma$ with characteristics of the disorder in the limiting cases. In general, the presence of impurities also renormalizes the chemical potential $\mu$. We will neglect that effect since it does not bring any new physics. In addition, this renormalization is small under experimentally achievable conditions. Then, instead of applying Eqs.~\eqref{dyson} and~\eqref{SelfEnergy}, we will use a scattering amplitude $\Gamma$ as a phenomenological parameter. It is very reasonable since this quantity can be measured in the experiment~\cite{Beidenkopf2011}. As a results, we obtain from Eqs.~\eqref{dyson} and~\eqref{green0} an expression for the impurity averaged Green's functions in the form
\begin{widetext}
	\begin{eqnarray}\label{greenF0}
	G^{\pm}=\frac{\mu\pm i\Gamma\!-\!v_F(k_x \sigma_y - k_y \sigma_x)\! - \! \lambda k_x (k_x^2-3k_y^2)\sigma_z\!}{(-\mu\pm i\Gamma)^2-\left\{(v_Fk_x)^2+(v_Fk_y)^2+\left[\lambda k_x(k_x^2-3k_y^2)\right]^2\right\}}\,.
	\end{eqnarray}
\end{widetext}

When the magnetic field is not too high, $B<\mu$, we can assume that $\mathbf{B}$ does not affect scattering significantly. In this case, we readily rewrite an expression for the impurity averaged Green's functions~\eqref{greenF0} of the Hamiltonian~\eqref{H0} with taking into account the Zeeman term:
\begin{widetext}
	\begin{eqnarray}\label{greenF}
	G^{\pm}=\frac{\mu\pm i\Gamma\!-\!v_F(k_x \sigma_y - k_y \sigma_x)\! - \! \lambda k_x (k_x^2-3k_y^2)\sigma_z\!-\!B_{\textrm{in}}\cos(\beta)\sigma_x - B_{\textrm{in}}\sin (\beta)\sigma_y-B_z\sigma_z}{(\mu\pm i\Gamma)^2-\left\{(v_Fk_x+B_{\textrm{in}}\sin\beta)^2+(v_Fk_y-B_{\textrm{in}}\cos\beta)^2+\left[\lambda k_x(k_x^2-3k_y^2)+B_z\right]^2\right\}}\,.
	\end{eqnarray}
\end{widetext}

\section*{Acknowledgements}

We acknowledge support from the Russian Scientific Foundation, Grant No 17-12-01544. RSA acknowledge the partial support by the Basis Foundation and ICFPM (MMK) of Education and Science of the Russian Federation, Grant No. 14Y26.31.0007.

\bibliographystyle{apsrevlong_no_issn_url}
\bibliography{bib_hw}

\end{document}